\documentclass[pra,twocolumn,preprintnumbers,amsmath,amssymb,nofootinbib,floatfix]{revtex4}

\usepackage{graphicx,bm}

\makeatletter
\def\graphicscale{\twocolumn@sw{0.3}{0.4}}
\def\graphicthreescale{\twocolumn@sw{0.3}{0.4}}

\begin{document}

\title{Critical mass renormalization in renormalized $\phi^4$ theories in two
   and three dimensions}

\author{Andrea Pelissetto$^1$ and Ettore Vicari$^2$} 

\address{$^1$ Dipartimento di Fisica dell'Universit\`a di Roma ``La Sapienza"
        and INFN, Sezione di Roma I, P.le Aldo Moro 2, I-00185 Roma, Italy}
\address{$^2$ Dipartimento di Fisica dell'Universit\`a di Pisa
        and INFN, Largo Pontecorvo 3, I-56127 Pisa, Italy}

\date{\today}

\begin{abstract}
We consider the O($N$)-symmetric $\phi^4$ theory in two and three
dimensions and determine the nonperturbative mass renormalization
needed to obtain the $\phi^4$ continuum theory. The required
nonperturbative information is obtained by resumming high-order
perturbative series in the massive renormalization scheme, taking into
account their Borel summability and the known large-order behavior of
the coefficients. The results are in good agreement with those
obtained in lattice calculations.
\end{abstract}


\maketitle



\section{Introduction}
\label{intro}

The $N$-vector $\phi^4$ theory for a field $\phi(x)$ with $N$
components, with Lagrangean
\begin{equation}
{\cal L} = {1\over 2} 
  \sum_\mu \partial_\mu \phi^2 + 
{1\over 2}\mu_0^2 \phi^2 + {1\over 4!}g \phi^4
\end{equation}
(where $\phi^2\equiv \phi\cdot \phi$ and $\phi^4\equiv (\phi^2)^2$),
is an important field theory model that can be used to describe a wide
variety of systems under critical conditions.  Because of the
ultraviolet divergences, a proper definition requires the introduction
of a regularization. We will use here the corresponding lattice theory
that has the advantage of being well defined also at the
nonperturbative level. On a regular $d$-dimensional cubic lattice the
action is given by
\begin{equation}
S = {1\over 2} \sum_{xy} J(x-y) \phi_x \phi_y +
    \sum_x \left( {1\over 2} \mu_0^2 \phi_x^2 +
       {1\over 4!} g \phi^4_x \right),
\label{modelS}
\end{equation}
where the fields $\phi_x$ are $N$-dimensional vectors.  We assume that
fields and $\mu_0^2$ are chosen so that the Fourier transform
$\tilde{J}(p)$ satisfies $\tilde{J}(p) = p^2 + O(p^4)$ for small
momenta. We make no other assumption on $J(x)$, so that model
(\ref{modelS}) represents the most general lattice model consistent
with the $\phi^4$ continuum theory.

In this paper we shall focus on the theory in $d=2$ and $d=3$. In this
case the model is superrenormalizable, which greatly simplifies the
determination of the continuum limit. Indeed, it is enough to perform
a (nonperturbative) mass renormalization.  If one defines a
renormalized mass $t = \mu^2_0 - \mu^2_{0c}$, the continuum limit is
obtained by considering $g\to 0$, $t\to 0$ at fixed $t g^{2/(4-d)}$,
where $d$ is the space dimension. In the statistical-mechanics
framework, $\mu^2_{0c}$ represents the value of the bare parameter
$\mu^2_0$ at which the statistical system undergoes a continuous
second-order transition. The determination of $\mu^2_{0c}$ is crucial,
as it representes a prerequisite in any study of the $\phi^4$ theory
in the continuum limit. Beside its field-theoretical interest,
$\mu^2_{0c}$ is also required in some calculations concerning dilute
relativistic and nonrelativistic Bose gases, in homogeneous conditions
and in the presence of trapping potentials
\cite{AM-01a,AM-01,AT-01a,AT-01b}.

The determination of $\mu^2_{0c}$ in the limit $g\to 0$ is not an easy
task as it represents a nonperturbative renormalization.  In two
dimensions it has been computed either by Monte Carlo simulations of
lattice models or by an analysis of the corresponding Hamiltonian
model defined in one dimension
\cite{LSL-01,Sugihara-04,SL-09,WW-12,MHO-13,RV-15,BDPG-15}.  In three
dimensions results, obtained by means of Monte Carlo simulations, are
available for $N=2$ \cite{AM-01}.  Here, we will perform a different
calculation, using the high-order perturbative series, computed in the
massive renormalization scheme, that provide the critical exponents
for the critical theory \cite{BNGM-77,MN-91}.  The resummation of
these perturbative series \cite{LZ-80}, taking into account their
Borel summability and the known large-order behavior of the
coefficients \cite{Large-order} allows us to obtain the
nonperturbative information that is needed to compute $\mu^2_{0c}$.
As we shall see, we obtain results for $d=2$ and $d=3$ with a
precision that is comparable with that obtained using state-of-the-art
numerical simulations of lattice models, confirming the accuracy of
resummed perturbation theory.

\section{Two dimensions}

Let us consider the generic $\phi^4$ model (\ref{modelS}) on a
two-dimensional square lattice. We only discuss the case $N=1$, as
only for this value of $N$ the model undergoes a standard transition
from a symmetric to a broken phase.  The superrenormalizability
properties of the theory allow us to predict
\begin{equation}
\mu_{0c}^2 = A g \ln g + B g + O(g^2 \ln^2 g),
\end{equation}
for $g \to 0$. The constant $A$ can be computed in perturbation theory
and does not depend on the regularization, while the constant $B$ is 
nonperturbative and regularization dependent. 

To determine $A$ and $B$ we consider the integrated bare two-point
correlation function
\begin{equation}
\chi = \sum_x \langle \phi_0 \phi_x\rangle.
\end{equation}
In \cite{BB-85,BBMN-87,PRV-99} it was shown that the combination
$\tilde{\chi} = \chi g$ is a regularization-independent function
$F_\chi(\tilde{t})$ of $\tilde{t} = t/g$, $t = \mu^2_0 - \mu^2_{0c}$
in the limit $g\to 0$, $t\to 0$ at fixed $\tilde{t}$. This limit,
which was called critical crossover limit \cite{PRV-98,PRV-99,PV-02},
corresponds to what we call continuum limit in the present context.
The quantity $\tilde{t}$ is the dimensionless renormalized mass: for
$\tilde{t} \to 0$ we obtain the critical massless regime, while for
$\tilde{t} \to \infty$ we recover the weak-coupling behavior.

The function $F_\chi(\tilde{t})$ is intrinsically nonperturbative.  In
\cite{PRV-99} it was determined by resumming the perturbative series
of renormalization-group invariant functions in the massive
renormalization scheme. Four-loop results were used \cite{BNGM-77},
taking explicitly into account \cite{LZ-80} the Borel summability of
the perturbative series and the large-order behavior of their
coefficients, determined by nonperturbative instanton calculations
\cite{Large-order}.  For $\tilde{t}\to\infty$ reference \cite{PRV-99}
obtained
\begin{equation}
F_\chi(\tilde{t}) = {1\over \tilde{t}} + 
   {1\over \tilde{t}^2} \left[
   {1\over 8\pi} \ln \left( {8 \pi \tilde{t}\over 3}\right) + 
   {3\over 8 \pi} + D_2 \right] + O(\tilde{t}^{-3} \ln^2 \tilde{t}),
\label{Fchi}
\end{equation}
where $D_2$ is a nonperturbative constant. Resummation of the
perturbative series gave $D_2 = -0.0524(2)$.

Let us now compute $\chi$ in the lattice model. At one loop we have
\begin{equation}
\chi = {1\over \mu_0^2} - {g\over 2\mu^4_0}
   \int {d^2p\over (2 \pi)^2} {1\over \tilde{J}(p) + \mu_0^2} + O(g^2).
\end{equation}
We now rewrite $\mu_0^2 = t + \mu_{0c}^2$ and expand all quantities
for $g\to 0$. Since
\begin{equation}
\int {d^2p\over (2 \pi)^2} {1\over \tilde{J}(p) + \mu_0^2} = 
   - {1\over 4\pi} \ln \mu_0^2 + K + O(\mu_0^2 \ln \mu_0^2),
\label{tadpole}
\end{equation}
where $K$ is a constant that depends on $J(x)$, we obtain
\begin{equation}
\tilde{\chi} \approx {1\over \tilde{t}} - 
    {\ln g\over \tilde{t}^2} \left(A - {1\over 8\pi}\right) + 
    {1\over 8\pi\tilde{t}^2} \ln \tilde{t} + 
    {1\over 2\tilde{t}^2} (K - 2 B),
\end{equation}
where we have replaced $t$ with $\tilde{t} = t/g$.  Comparison with
(\ref{Fchi}) gives
\begin{eqnarray}
 A &=& {1\over 8\pi}, \nonumber \\
 B &=& - {3\over 8\pi} - {1\over 8\pi} \ln {8\pi \over 3} - {K \over 2} - D_2.
\label{stimeAB}
\end{eqnarray}
The constant $B$ depends on the regularization through the constant
$K$, as expected for a bare mass term.

To compare with the results reviewed in \cite{BDPG-15}, let us
introduce the perturbatively renormalized mass $\mu^2$ defined by
\begin{equation}
 \mu_0^2 = \mu^2 - {g\over 2}
  \int {d^2p\over (2 \pi)^2} {1\over \tilde{J}(p) + \mu^2}.
\end{equation}
We wish now to compute the critical value $\mu_c^2$ that corresponds
to $\mu_{0c}^2$. We find $\mu_c^2 = C g$, where $C$ is independent of
the regularization. Using (\ref{stimeAB}) and (\ref{tadpole}) we find
that $C$ satisfies the equation
\begin{equation}
C + {1\over 8\pi} \ln C = - D_2 - {3\over 8\pi} - 
    {1\over 8\pi} \ln {8\pi\over 3}.
\end{equation}
Note that $K$ cancels out, proving the regularization independence of
$C$.  Solving this equation we obtain
\begin{equation}
 C = 0.01515(6),
\end{equation}
where the reported error is related to the uncertainty on
$D_2$. Taking into account the different normalization of the coupling
constant, we obtain for the quantity $f_0$ defined in \cite{BDPG-15}
\begin{equation}
f_0 = {1\over 6 C} = 11.00(4).
\label{f0}
\end{equation}
The quantity $f_0$ has been also computed by means of other techniques
\cite{LSL-01,Sugihara-04,SL-09,WW-12,MHO-13,RV-15,BDPG-15}.
References \cite{SL-09,WW-12,BDPG-15} use Monte Carlo methods, while
\cite{MHO-13,RV-15} consider the Hamiltonian quantum formulation in
one dimension.  Results are reviewed in \cite{BDPG-15}. The most
recent estimates ($f_0 = 10.92(13), 11.06(2), 11.88(56), 11.15(9)$ of
\cite{WW-12}, \cite{MHO-13}, \cite{RV-15}, and \cite{BDPG-15},
respectively) are all in good agreement with our result. Note also
that the error on the estimate (\ref{f0}) is comparable with those
obtained using state-of-the art numerical algorithms, confirming the
accuracy of resummed perturbation theory.

\section{Three dimensions}

Analogous considerations apply to three dimensions. Using the results
of \cite{PRV-99} we are now going to compute the nonperturbative mass
renormalization for the three-dimensional theory.  We start from the
two-loop expansion of $\chi$ in powers of the bare coupling constant
$g$:
\begin{eqnarray}
\chi^{-1} &=& \mu_0^2 + {N+2\over 6} g T_1(\mu_0^2) - 
       {N+2\over 18} g^2 T_3(\mu_0^2) 
\nonumber \\
    && - 
       \left({N+2\over 6}\right)^2 g^2 T_1(\mu_0^2) T_2(\mu_0^2),
\end{eqnarray}
where 
\begin{eqnarray}
&&T_1(m^2) = \int {d^3p\over (2\pi)^3} {1\over \Delta(p)} ,
\nonumber \\
&&T_2(m^2) = \int {d^3p\over (2\pi)^3} {1\over \Delta(p)^2} ,
\nonumber \\
&&T_3(m^2) = \int {d^3p\over (2\pi)^3} {d^3q\over (2\pi)^3}
   {1\over \Delta(p)\Delta(q)\Delta(p+q)}.
\end{eqnarray}
with $\Delta(p) = \tilde{J}(p) + m^2$.
The continuum limit is obtained by tuning $\mu_0^2$ to the critical value
$\mu_{0c}^2$. More precisely, $\tilde{\chi} = \chi g^2$ becomes a 
universal function of the renormalized mass $\tilde{t} = t/g^2$, where 
$t = \mu^2_0 - \mu_{0c}^2$, for $t\to 0$, $g \to 0$ at fixed $\tilde{t}$.

Let us now proceed as in two dimensions. We expand 
\begin{equation}
\mu_{0c}^2 = Ag + B g^2 \ln g + C g^2,
\end{equation}
rewrite $\mu_0^2 = t + \mu_{0c}^2$ and expand in powers of $g$. 
Now, for $m\to 0$ we have 
\begin{eqnarray}
T_1(m^2) &=& T_1(0) - {m\over 4\pi} + K_1 m^2 + O(m^3) \nonumber \\
T_2(m^2) &=& {1\over 8\pi m} - K_1 + O(m) \nonumber \\
T_3(m^2) &=& -{1\over 32\pi^2} \ln m^2  + K_2  + O(m).
\end{eqnarray}
Using these expressions, for $t,g\to 0$ we obtain
the expansion 
\begin{eqnarray}
&& \tilde{\chi}^{-1} = 
   g^{-1} \left(A + {N+2\over 6} T_1(0)\right) 
     \left(1 - {N+2\over 48\pi}  {1\over \tilde{t}^{1/2}}\right) 
   \nonumber \\ 
&& \quad   + {N+2\over 6} K_1  \left(A + {N+2\over 6} T_1(0)\right)
\nonumber \\  
 &&  \quad + \ln g \left(B + {N+2\over 288\pi^2}\right) - {N+2\over 18} K_2
\label{expansion-chi-3d}
 \\
   && \quad + 
      \tilde{t} -
    {N+2\over 24 \pi} \sqrt{\tilde{t}} + 
    {N+2\over 576 \pi^2} \ln \tilde{t} + C
    + {(N+2)^2\over 1152 \pi^2}.
\nonumber 
\end{eqnarray}
Cancellation of the terms of order $1/g$ and $\ln g$ gives
\begin{equation}
A = - {N+2\over 6} T_1(0) \qquad 
B = - {N+2\over 288 \pi^2}.
\end{equation}
Finally, we compare (\ref{expansion-chi-3d}) with the expression given 
in \cite{PRV-99}:
\begin{equation}
\tilde{\chi} = {1\over \tilde{t}} + 
    {N+2\over 24 \pi} \tilde{t}^{-3/2} - 
    {N+2\over 576 \pi^2} {\ln \tilde{t}\over \tilde{t}^2} + 
    {E\over \tilde{t}^2} ,
\end{equation}
where $E$ was computed by using resummed perturbation theory (in this 
case seven-loop expansions are available \cite{MN-91}). Comparing the two 
expressions we obtain 
\begin{equation}
C = - E + {N+2\over 18} K_2  + {(N+2)^2\over 1152 \pi^2}.
\label{eqC}
\end{equation}
This expression shows that $C$ is regularization dependent. 
To determine its explicit value, we compute the constant $E$ using 
the results of \cite{PRV-99}, obtaining
\begin{eqnarray}
E = -0.002504(6)   && \qquad N = 1\\
E = -0.002885(5)   && \qquad N = 2\\
E = -0.003042(3)   && \qquad N = 3,
\end{eqnarray}
while for $N\to \infty$, we have $E \approx N^2/(1152 \pi^2) + O(N)$.
Correspondingly, if $\widehat{C} = C - {N+2\over 18} K_2$, we have
\begin{equation}
\begin{array}{ll} 
\widehat{C} = 0.003296(6)  & \qquad N = 1\\
\widehat{C} = 0.004292(5)  & \qquad N = 2\\
\widehat{C} = 0.005241(3)  & \qquad N = 3.
\end{array}
\end{equation}
For $N\to \infty$, the terms of order $N^2$ cancel, so that $\widehat{C}$ 
is of order $N$. 

It is interesting to extend the calculation to the continuum model in
dimensional regularization, to compare with the result of \cite{AM-01}
for $N=2$. In this scheme $T_1(0)= 0$. Regularizing $T_2$ in $d =
3-\epsilon$, and renormalizing it by minimal subtraction, we obtain
(we use the results of \cite{BFT-93,DT-93})
\begin{equation}
K_2 = {1\over 16 \pi^2} \ln \overline{\mu} + K_{20},
\end{equation}
where $\overline{\mu}$ is the renormalization scale in the 
$\overline{\rm MS}$ scheme ($\overline{\mu} = \sqrt{4\pi} \mu e^{-\gamma_E/2}$)
and $K_{20} = (1 - 2 \ln 3)/(32\pi^2) \approx -0.00379076$. 
We can thus rewrite 
\begin{equation}
{\mu_{0c}^2\over g^2} = - {N+2\over 288 \pi^2} \ln {g\over \overline{\mu}}
+ {N+2\over 18} K_{20}  + \widehat{C}
\end{equation}
We can compare this result with that reported in \cite{AM-01}. For
$N=2$ we obtain ${\mu_{0c}^2/ g^2} = 0.001904(5)$ for $N = 2$ and
$g/\overline{\mu} = 3$, to be compared with the numerical estimate
$0.001920(2)$ of \cite{AM-01}.  The two results are close, although
they do not properly agree within errors (in any case, the difference
is still acceptable being of the order of twice the sum of the error
bars).

\section{Conclusions}

We considered the $O(N)$ invariant $\phi^4$ theory in two and three
dimensions and determined the nonperturbative mass renormalization
$\mu_{0c}^2$ one must perform to obtain the continuum limit of the
model.  Such a quantity is also relevant in the context of dilute
relativistic and nonrelativistic Bose gases
\cite{AM-01a,AM-01,AT-01a,AT-01b}. In two dimensions there are several
computations
\cite{LSL-01,Sugihara-04,SL-09,WW-12,MHO-13,RV-15,BDPG-15} of
$\mu_{0c}^2$ for $N=1$. In three dimensions it has been determined
\cite{AM-01} for $N=2$, the relevant case for Bose gases.

We computed the mass-renormalization constant $\mu_{0c}^2$ for a
generic lattice model in $d=2$ and $d=3$. The necessary
nonperturbative information was taken from Ref.~\cite{PRV-99}, where
several nonperturbative quantities where computed in the continuum
limit (in that context the continuum limit was named critical
crossover limit) as a function of the dimensionless renormalized
mass. They were estimated by resumming the perturbative series in the
massive renormalization scheme (four-loop \cite{BNGM-77} and
seven-loop \cite{BNGM-77,MN-91} results are available in $d=2$ and
$d=3$, respectively), taking explicitly into account \cite{LZ-80} the
Borel summability of the perturbative series and the large-order
behavior of their coefficients, determined by nonperturbative
instanton calculations \cite{Large-order}.

The results we obtain are in good agreement with present-day
state-of-the-art numerical determinations, confirming the accuracy of
resummed perturbation theory.


\begin{thebibliography}{99}

\bibitem{AM-01a}
P. Arnold and G.D. Moore,
Phys. Rev. Lett. {\bf 87} (2001) 120401, arXiv:cond-mat/0103228.

\bibitem{AM-01}
P. Arnold and G.D. Moore,
Phys. Rev. E {\bf 64} (2001) 066113, arXiv:cond-mat/0103227;
(erratum) Phys. Rev. E {\bf 68} (2003) 049902.

\bibitem{AT-01a}
P. Arnold and S. Tkachenko,
Phys. Rev. D {\bf 64} (2001) 105018.

\bibitem{AT-01b}
P. Arnold and S. Tom\'a{\v s}ik, 
Phys. Rev. A {\bf 64} (2001) 053609,  arXiv:cond-mat/0105147;
P. Arnold, G.D. Moore, B. Tom\'a{\v s}ik,
Phys. Rev. A {\bf 65} (2002) 013606, cond-mat/0107124.

\bibitem{LSL-01}
D. Lee, N. Salwen and D. Lee, 
Phys. Lett. B 503 (2001) 223,
arXiv:hep-th/0002251.

\bibitem{Sugihara-04}
T. Sugihara, JHEP 0405007 (2004),
arXiv:hep-lat/0403008.

\bibitem{SL-09}
D. Schaich and W. Loinaz, 
Phys. Rev. D {\bf 79} (2009) 056008, arXiv:0902.0045.

\bibitem{WW-12}
C. Wozar and A. Wipf, 
Ann. Phys. {\bf 327} (2012) 774, arXiv:1107.3324.

\bibitem{MHO-13}
A. Milsted, J. Haegeman, and T.J.  Osborne,
Phys. Rev. D 88, 085030 (2013), arXiv:1302.5582.

\bibitem{RV-15}
S. Rychkov and L.G. Vitale, 
Phys. Rev. D 91, 085011 (2015), arXiv:1412.3460.

\bibitem{BDPG-15}
P. Bosetti, B. De Palma and M. Guagnelli,
arXiv:1506.08587.

\bibitem{BNGM-77}
G.A. Baker Jr., B.G. Nickel, M.S. Green and D.I. Meiron, 
Phys. Rev. Lett. {\bf 36} (1977) 1351;
G.A. Baker Jr., B.G. Nickel and D.I. Meiron, Phys. Rev. B 17 (1978) 1365.

\bibitem{MN-91}
D.B. Murray and B.G. Nickel, Revised estimates for critical exponents for the
continuum n-vector model in 3 dimensions, unpublished Guelph University report 
(1991).

\bibitem{LZ-80}
J.C. Le Guillou and J. Zinn-Justin, Phys. Rev. B {\bf 21} (1980) 3976.

\bibitem{Large-order}
L.N. Lipatov, Zh. Eksp. Teor. Fiz. {\bf 72} (1977) 411 
[JETP {\bf 45} (1977) 216];
E.  Br\'ezin, J.C. Le Guillou and J. Zinn-Justin, 
Phys. Rev. D {\bf 15} (1977) 1544.

\bibitem{BB-85} 
C. Bagnuls and C. Bervillier, Phys. Rev. B {\bf 32} (1985) 7209.

\bibitem{BBMN-87}
C. Bagnuls, C. Bervillier, D.I. Meiron and B.G. Nickel, 
Phys. Rev. B {\bf 35} (1987) 3585.

\bibitem{PRV-99} 
A. Pelissetto, P. Rossi and E. Vicari, 
Nucl. Phys. B {\bf 554} [FS] (1999) 552, arXiv:cond-mat/9903410.

\bibitem{PRV-98}
A. Pelissetto, P. Rossi and E. Vicari, 
Phys. Rev. E {\bf 58} (1998) 7146, arXiv:cond-mat/9804264.

\bibitem{PV-02}
A. Pelissetto and E. Vicari,
Phys. Rep. {\bf 368} (2002) 549.

\bibitem{BFT-93}
D.J. Broadhurst, J. Fleischer and O.V. Tarasov, Z. Phys. C {\bf 60} (1993) 287,
arXiv:hep-ph/9304303;

\bibitem{DT-93}
A.I. Davydychev and J.B. Tausk, Nucl. Phys. B {\bf 397} (1993) 123.


\end{thebibliography}
\end{document}